# 28 GHz NLOS Channel Measurements Revealing Low Path Loss and High Angular Spread in Container Ports

Andrea Bedin *Graduate Student Member, IEEE*, Dmitry Chizhik *Fellow, IEEE*, Jinfeng Du *Member, IEEE*, Martti Moisio *Member, IEEE*, Karthik Upadhya *Member, IEEE*, Reinaldo Valenzuela *Fellow, IEEE*, and Mikko A. Uusitalo *Senior Member, IEEE*

*Abstract*—This paper presents results from a comprehensive measurement campaign conducted at 28 GHz inside a container canyon within a commercial port environment. The measurements are performed at various points inside the container canyon, considering two types of container stacking and two different Transmitter (TX) locations, using a narrowband channel sounder equipped with a rotating horn antenna. The measurements are used to evaluate the azimuthal spectrum and spatial correlation, as well as the impact of a vehicle inside a canyon on these parameters. Further, the measurement data is utilized to validate a simulation setup from which the path loss and the elevation spectrum inside the canyon is obtained. Lastly, a propagation model inside the canyon is hypothesized and shown to be consistent with the measurements. The analysis show a low path loss compared to free space, as well as a high angular spread and short spatial correlation.

## I. INTRODUCTION

Communication at Millimeter wave (mmWave) frequencies is a well studied topic [1], [2], [3] that holds great potential for industrial automation, given the large bandwidths and high outdoor-to-indoor pathloss that minimizes interference from outdoor access points in factories. In addition, the large bandwidths enable low latencies and high throughputs crucial for machine-to-machine communication, especially when carrying bandwidth-intensive traffic, such as video, from several User Equipments (UEs) to the edge cloud [4], [5].

Port automation is an important subset of industrial automation and is a growing trend in the maritime industry that aims to enhance efficiency, reduce costs, improve safety, and increase productivity throughout the entire port ecosystem [6].

While mmWaves offer large bandwidths, their propagation characteristics are quite different from sub-6 GHz. Signal propagation at mmWaves is characterized by limited diffraction around objects resulting in an increased pathloss and blocking when compared to sub-6 GHz frequencies [5]. Beamforming and Multiple-Input Multiple-Output (MIMO) are also essential for meeting the link budget, however, most modern beamforming [7], [8], [9] and Channel State Information (CSI) acquisition [10], [11] methods relies on a sparse channel like the one observed in a mmWave indoor setting [12], [13]. In contrast, industrial settings show a rich multipath [14], [15], [16]. Further, port automation and mmWave communication form an interesting combination since the challenging propagation environment, which is open-air and characterized by canyons of stacked metallic containers, can potentially block mmWaves from reaching UEs such as forklifts that may be moving in between the container stacks, making it important to characterize the propagation environment to determine coverage, beam-training intervals, and beam patterns.

Measurement campaigns are essential for characterizing the environment and for building propagation models for system-level simulations and the characterization has to be redone for different types of factory environments [17]. Several measurement campaigns at mmWave frequencies are available in the literature for a variety of environments [14], [15], [16]. In [14] and [15], measurements were collected with a narrow-band sounder at 28 GHz in an indoor factory which was then used to characterize the dependence of path-gain on range and the effective antenna gains degraded by scattering. In [16], measurements inside a factory were obtained with a wide-band channel sounder and parameters such as path loss, delay, angular spread, and cross-polarization ratio were evaluated. Measurements at 60 GHz in a factory environment were obtained in [18]. [19] proposes a path-loss model for mmWave propagation in an underground mine tunnel based on simulations and measurements. Despite there being measurement campaigns and models for a wide variety of scenarios, a characterization of mmWave communication in a commercial port environment has not been done, to the best of our knowledge.

In this paper, we describe a measurement campaign carried out at 28 GHz in a commercial port environment, specifically inside a container canyon. The measurements were performed with a narrow-band channel sounder at different points inside the canyon for two types of container stacking and two different Transmitter (TX) locations. At each measurement point

A. Bedin (Corresponding author, andrea.bedin.2@studenti.unipd.it), M. Moisio (martti.moisio@nokia-bell-labs.com), K. Upadhya (karthik.upadhya@nokia-bell-labs.com) and M. Uusitalo (mikko.uusitalo@nokia-bell-labs.com) are with Nokia Bell Labs Finland. A. Bedin is also with the University of Padova, Department of Information Engineering. D. Chizhik (dmitry.chizhik@nokia-bell-labs.com), J. Du (jinfeng.du@nokia-bell-labs.com) and R. Valenzuela (reinaldo.valenzuela@nokia-bell-labs.com) are with Nokia Bell Labs US. This work has received funding from the European Union's EU Framework Programme for Research and Innovation Horizon 2020 under Grant Agreement No 861222.

| Section | 1 | 2 | 3 | 4 | 5 | 6 |
|---|---|---|---|---|---|---|
| Row 1 | 10m | 7.5m | 5m | 5m | 7.5m | 5m |
| Row 2 | 5m | 5m | 5m | 7.5m | 7.5m | 7.5m |

TABLE I: Height of the canyon in the nonuniform case.

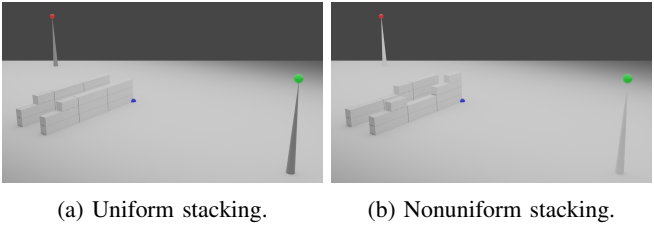

(a) Uniform stacking.  (b) Nonuniform stacking.

Fig. 1: Measurement environment. Transmitter 1 (TX1) and Transmitter 2 (TX2)depicted in green and red respectively.

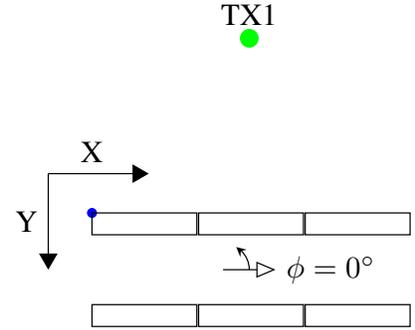

Fig. 2: Coordinate reference (distance not to scale).

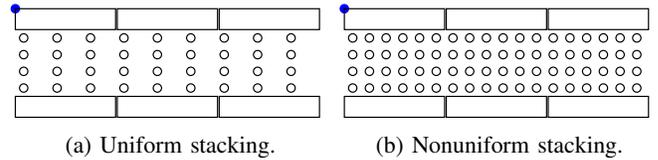

(a) Uniform stacking.  (b) Nonuniform stacking.

Fig. 3: Coarse measurements map.

inside the canyon, the angular distribution of the receive power is obtained from the sounder, which in turn is used to evaluate the angular spread, and spatial correlation. Measurements were also obtained with a vehicle present at the entrance of the container canyon to study the impact of the vehicle on these parameters. The measurements are also used to validate a simulation setup, which is then used to obtain the path loss and elevation spectrum inside the canyon. A mathematical model for propagation is also proposed for this environment.

In Section II, we describe the measurement setup and in Section III, we discuss the angular distribution of energy in the azimuthal direction inside the canyon to understand the gains obtained with beamforming. In Section IV we analyze the spatial correlation of the pathloss to determine how often the UE would have to perform beamtraining. We then look at the impact of a large vehicle present in the canyon on the azimuthal distribution of energy, and in Section VI, a pathloss model is obtained by fitting a model to the measurements. Section VII presents a model for the received power inside the canyon and validates it with the measurements. Section VIII studies the elevation spectrum and Section IX concludes the paper.

## II. MEASUREMENT SETUP

The measurements were conducted with the equipment described in [12]. In particular we used a narrowband TX operating at 28 GHz with an output power of 22 dBm and an omnidirectional antenna, whereas the Receiver (RX) is equipped with a $10°$ Half Power Beam Width (HPBW) horn antenna. The RX antenna is rotated in the azimuthal plane at a speed of 120 rpm and at each measurement point we collected 10 s of power measurements, corresponding to 20 full rotations of the antenna. The measurement environment, depicted in Fig. 1, was purposely built to resemble a commercial port and consists of a set of shipping containers arranged in the shape of a canyon that is 36 m long with an internal width of 8 m.

The height of the container canyon has been changed during the campaign between two different configurations. In the first configuration (Fig. 1a), hereafter referred to as the uniform configuration, the height of the canyon was of 3 containers (roughly 7.5 m) for almost all the length of the canyon, with the exception of the last 6 m which were 2 containers (5 m) high. The second configuration (Fig. 1b), referred to as the non-uniform configuration has a varying height. The height of each section of the canyon is listed in Table I.

In order to define the positions of the TXs and RXs, we define as a reference point one of the corners of the canyon (depicted in blue in Fig. 1). We denote the direction parallel to the canyon as "X" direction and the one orthogonal the canyon as "Y" direction, as depicted in Fig. 2. When not otherwise specified, all the distances are measured from the reference corner. We used two different TX positions: TX1 position, depicted in green was placed at 18.8 m from the reference corner in the X direction, and was mounted on a rail crane that could move from 63 m to 113 m in the Y direction. The height of TX1 from the ground was 23 m. TX2 was placed on a pole that is at 18.85 m in the X direction and 60.5 m in the Y direction. The height of TX2 was of 22 m from the ground.

The RX, representing the UE, was placed in a number of different positions, based on several predefined maps. Throughout the whole campaign, the reference direction corresponding to the $0°$ angle has been the one depicted in Fig. 2, where the big arrow represents the $0°$ direction and the small arrow represents the positive angle direction.

The first set of measurements is a coarse spatial sampling, and the measurement points for this experiment are depicted in Fig. 3 as black circles. In both the uniform and nonuniform stacking case, the measurements are on 4 lines at a distance of 3.5, 5.5, 7, 5 and 9.5 meters from the reference point in the Y direction, and starts at 1 m in the X direction. In the uniform case the spacing between the measurements in the X direction is of 4 meters, whereas in the nonuniform case it is 2 meters. The reason behind this difference in measurement step-size is that in the uniform case, the variation in the channel statistics inside the canyon in the X direction can be expected to be small owing to the uniformity of the canyon height. However, with non-uniform canyon height, one can expect the path-loss to change more often in the X direction necessitating a smaller measurement step.

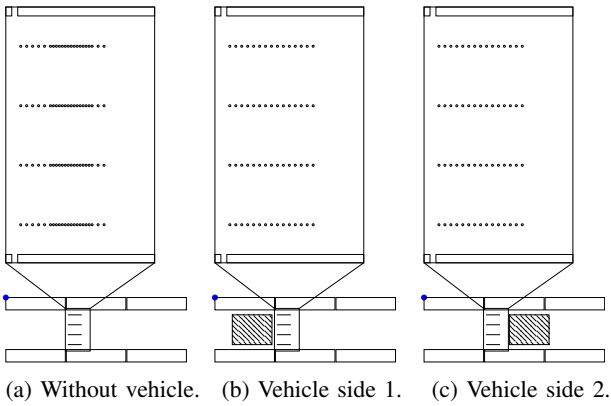

(a) Without vehicle.  (b) Vehicle side 1.  (c) Vehicle side 2.

Fig. 4: Dense measurements map.

For the uniform case, the measurement has been repeated for different positions of TX1, ranging from 63.5 m to 113.5 m in steps of 10 m, and from TX2. For the nonuniform case, the crane was moved in steps of 20 m due to the higher number of RX points.

A second set of measurements was conducted on a denser grid depicted in Fig. 4. This measurement has two goals:

1) Studying the spatial correlation of the channel gain.
2) Determining the effect of the presence of a vehicle in the canyon.

The measurement points are on the same lines of the coarse measurement in the Y direction, whereas in the X direction the measurement points start at 12.5 m and ends at 15.3 m with a spacing of 20cm. Moreover, in the case where the vehicle is absent, from 13.5 m to 14.9 m the spacing is decreased to 10cm to obtain a finer sampling of the spatial correlation. These measurements have been performed for TX2 with nonuniform stacking, and for TX1 at 63 m for uniform stacking. Further TX positions were planned for this measurement, but due to weather and time constraints, it was not possible to execute them.

For each of the RX's position $(x, y)$ we obtain a channel gain measurement from transmitter $tx$, when the receive antenna is oriented in direction $\phi$, that we call $R(tx, x, y, \phi)$. Note that, in this case, we measure the channel gain between the TX and RX antenna ports, thus the resulting value includes the antenna gains.[1] Here, $\phi$ is the azimuthal direction as can be seen in Fig. 2. In this notation, the TX locations are denoted as $tx \in \{TX1_d, d \in \{63, 73, ..., 113\}; TX2\}$ where $d$ represents the Y component of the distance of the crane

[1] Sometimes referred to as coupling gain.

from the reference point.

We also define the received power in dB as:

$$R^{dB}(tx, x, y, \phi) = 10 \log_{10}(R(tx, x, y, \phi)) \quad (1)$$

and the normalized angular spectrum as

$$\hat{R}^{dB}(tx, x, y, \phi) = R^{dB}(tx, x, y, \phi) \\ - 10 \log_{10}\left(\frac{1}{2\pi} \int_0^{2\pi} R(tx, x, y, \phi) d\phi\right). \quad (2)$$

## III. ANGULAR SPREAD

In this section, we discuss how the received energy is distributed in the azimuthal direction. In particular, in Figs. 5 and 6, we can observe the histogram and average value of the normalized angular spectrum over a set of TX and RX positions. More precisely, the black line represents the mean

$$\mu(tx, \phi) = \frac{1}{|\mathcal{X}||\mathcal{Y}|} \sum_{x \in \mathcal{X}} \sum_{y \in \mathcal{Y}} R(tx, x, y, \phi) \quad (3)$$

the color represents the histogram of $R(tx, x, y, \phi)$ over all possible values of $x \in \mathcal{X}$ and $y \in \mathcal{Y}$, given $tx$ and $\phi$. Here, $\mathcal{X}$ and $\mathcal{Y}$ correspond to the $x$ and $y$ coordinates of the measurement points in Figs. 3a and 3b. In particular, Fig. 5a shows the statistics of the angular spectrum with TX1 positioned at a Y distance of $d \in \{63, 73, ..., 113\}$ m and Fig. 5b for TX2, with the RX is in all the positions depicted in Fig. 3a and uniform stacking. Fig. 6a depicts the statistics of the angular spectrum for TX1 at $d \in \{63, 83, 103\}$ m and Fig. 6b for TX2, with the RX in all the positions depicted in Fig. 3b and nonuniform stacking.

In the figures, we can observe that in the vast majority of cases, the channel gain measurements are within 10 dB from the average gain for all angles, This suggests that the energy is reaching the RX rather uniformly from all directions. We can also notice that the channel gain is slightly lower for the $0°$ and $180°$ directions in Figw. 5a and 6a. These two directions correspond to the horn antenna at the RX pointed parallel to the canyon. This result can be explained by the fact that since the RX is in Non Line of Sight (NLoS), we therefore expect that the energy is scattered or reflected by some object before reaching the RX. Indeed, in the $0°$ and $180°$ directions there is no object that can cause scattering or reflections.

In Figs. 5b and 6b, where the TX is in position TX2, we can instead observe how the reduction in gain happens mainly at $0°$. The $180°$ direction, which is facing towards the TX, shows a lower reduction in gain. Despite still not having Line of Sight (LoS) to the TX, the lower reduction in such direction can be explained by the guiding effect of the canyon. In other words, TX2 is placed at an angle of around $135°$ to the canyon close to the edge when compared to TX1 which is orthogonal to the canyon direction and located halfway along the canyon. As a consequence of this, some energy entering from the side of the canyon close to TX2 results in lower channel gain reduction at $180°$.

In Fig. 7a, we depict the distribution of $\hat{R}^{dB}(tx, x, y, \phi)$ computed over all directions $\phi \in [0, 2\pi)$ in comparison to

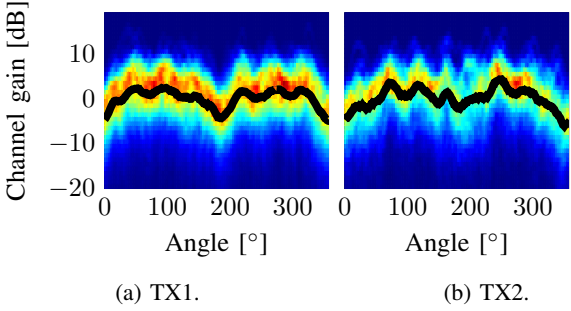

(a) TX1.  (b) TX2.

Fig. 5: Angular spectrum with uniform stacking.

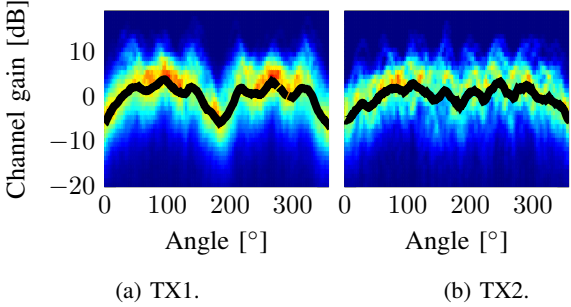

(a) TX1.  (b) TX2.

Fig. 6: Angular spectrum with nonuniform stacking.

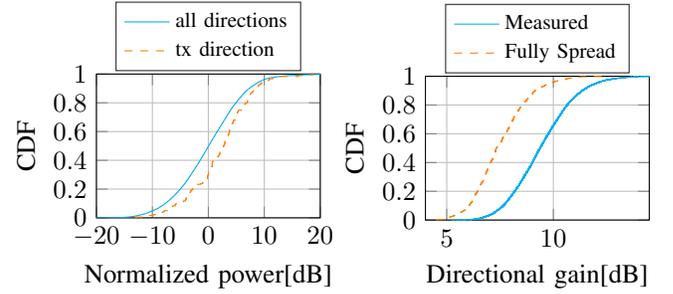

(a) Distribution of the normalized channel gain over all directions and the TX direction.

(b) Azimuth directional gain.

Fig. 7: Ineffectiveness of the beamforming in a port environment.

the distribution of $\hat{R}^{dB}(tx, x, y, \phi_{tx}(tx, x, y))$ where, given the location of the TX $(x_{tx}, y_{tx})$,

$$\phi_{tx}(tx, x, y) = \text{atan2}(y - y_{tx}, x_{tx} - x) \quad (4)$$

is the direction of the transmitter $tx$ when the RX is placed in position $(x, y)$. As can be observed in the plot, the two distributions are very close, with a difference of at most 2.8dB. This shows that having a narrow beam directly pointing towards the TX will not provide any significant benefit in the real world.

Furthermore, Fig. 7b shows the distribution of the azimuth gain $\max_\phi(\hat{R}^{dB}(tx, x, y, \phi))$ over all TX-RX locations, compared to the simulated distribution for a fully spread channel. Here we can also see a gap around 2dB, further confirming that azimuthal beamforming is not effective.

## IV. SPATIAL CORRELATION

In this section, we analyze the spatial correlation of $R^{dB}(tx, x_k, y, \phi)$ for a fixed value of $\phi$. This represents the channel correlation that would be observed by a user with a fixed beam, and therefore provides information on how often the beam should be updated.

To perform this analysis, we use the dense measurement map represented in Fig. 4a. In particular, we use the data collected during the dense sampling, sampled with a spatial period of 10 cm. We first define the set sequence of X positions corresponding to the 10 cm sampling as **x** such that $\{x_1 = 13.5, x_2 = 13.6, ..., x_{15} = 14.9\}$. We also define the average channel gain for a specific Y position and angle

$$m(tx, y, \phi) = \frac{1}{15} \sum_{k=1}^{15} R^{dB}(tx, x_k, y, \phi) \quad (5)$$

and the relative zero mean channel gain

$$R_0^{dB}(tx, x, y, \phi) = R^{dB}(tx, x, y, \phi) - m(tx, y, \phi)). \quad (6)$$

With these definition, we can compute the spatial autocorrelation of the channel gain[2] in the $x$ direction as

$$r(tx, x_k, y, \phi) = \sum_{j=1}^{15-k} R_0^{dB}(tx, x_k, y, \phi) R_0^{dB}(tx, x_{(k+j)}, y, \phi) \quad (7)$$

and average it over all the angles and Y positions to obtain:

$$r(tx, x_k) = \frac{1}{|\mathcal{Y}|} \sum_{y \in \mathcal{Y}} \frac{1}{2\pi} \int_0^{2\pi} r(tx, x_k, y, \phi) d\phi \quad (8)$$

In Fig. 8, we can see the plot of $r(tx, x_k)$ for the three measured cases. In the plot, we can clearly see that already after 10 cm the channel gain is completely uncorrelated. This in turns means that the "best direction" remains such for a very short time. This short spatial correlation is consistent with the hypotheses that the energy is received uniformly from all directions, and the best beamforming direction is purely an artifact of fading. It should be noted that, in this case, the best direction would also be frequency selective, further degrading the beamforming gain.

## V. VEHICLE IMPACT

In this section, we investigate how the presence of a large vehicle inside the canyon, at a few meters from the measurement points, impacts the channel gain. In particular, we define as $R_{v1}^{dB}(tx, x, y, \phi)$ the measured channel gain with the vehicle on side 1 and as $R_{v2}^{dB}(tx, x, y, \phi)$ the measured channel gain with the vehicle on side 2. With this definition, we can compute the received channel gain difference with the vehicle in the two positions as:

$$\Delta_1(tx, x, y, \phi) = R^{dB}(tx, x, y, \phi) - R_{v1}^{dB}(tx, x, y, \phi) \quad (9)$$
$$\Delta_2(tx, x, y, \phi) = R^{dB}(tx, x, y, \phi) - R_{v2}^{dB}(tx, x, y, \phi) \quad (10)$$

---

[2]The channel gain is expressed in dB, as such scale is used for most communication procedures, such as MCS selection and beam refinement. Thus it is more representative of how often these procedures needs to be performed

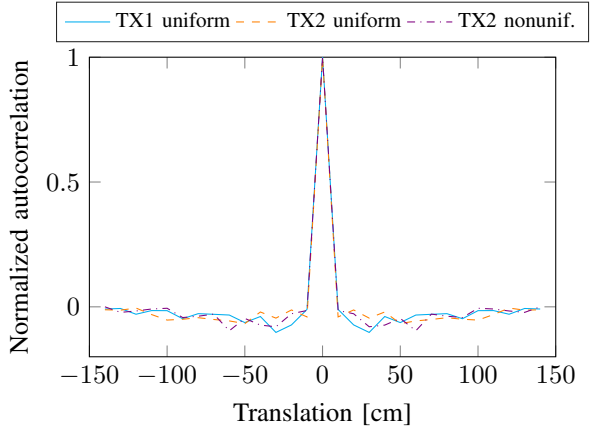

Fig. 8: Spatial correlation.

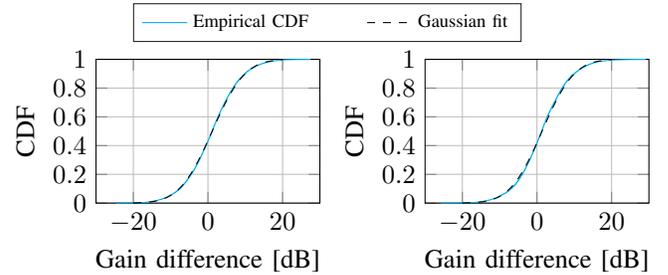

| | $\mu$ | $\sigma$ |
|---|---|---|
| Vehicle position 1 | 1.13 | 6.91 |
| Vehicle position 2 | 1.37 | 6.77 |

TABLE II: Channel gain difference Gaussian approximation parameters for the uniform stacking and TX1 at 63m.

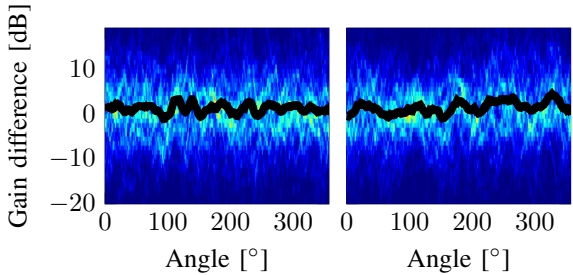

(a) Vehicle position 1.   (b) Vehicle position 2.

Fig. 9: Statistics of the vehicle effect with uniform stacking and TX1 at 63m.

(a) Vehicle position 1.   (b) Vehicle position 2.

Fig. 10: Angle independent statistics of the vehicle effect with uniform stacking and TX1 at 63m.

In Fig. 9 we can observe the statistics of $\Delta_1(tx,x,y,\phi)$ and $\Delta_2(tx,x,y,\phi)$ as a function of the angle over all RX positions, when the TX is in position TX1 at $63m$ in the y direction from the reference point, and the container stacking is uniform. Here the color represents the histogram of the channel gain difference and the black line represents its mean. We can observe that the difference is independent from the angle and with a mean around 0. This suggests that the presence of the vehicle is only impacting the local value of the channel gain, but not the overall statistics.

The angle independent Cumulative Distribution Function (CDF) of the channel gain difference, computed by aggregating the data over all angles, is depicted in Fig. 10, along with a fitted Gaussian distribution. Here we can observe that the CDF matches almost exactly the one of the Gaussian, and the mean is very close to 0, confirming that the impact of the vehicle on the statistics of the channel is negligible. In particular, the parameters for the fitted Gaussian distributions are listed in Tab. II

In Figs. 11 and 12 we can observe the same plots for the nonuniform stacking and the TX in positions TX2. Indeed, also in this case the effect on the average channel gain is limited, and the difference is approximately Gaussian with the parameters listed in Tab. III.

## VI. PATH LOSS

In this section, we discuss the path loss of the measured links and its impact on the communication system. In this scenario, the use of a narrow HPBW horn antenna rotated only in azimuth for these measurements poses a challenge in estimating the performance of a communication system with different antenna patterns. It, in fact, does not provide any information on the elevation angle at which the energy is received. This issue is particularly relevant in this scenario, where the energy is propagating from the top of a tall structure down to the user and it is very likely that the energy will reach the user from above.

In order to obtain reliable channel information for different antenna patterns, we used the data to validate a simulation setup. We recreated the environments and devices in CST microwave studio[3] and performed the simulation with the asymptotic physical optics solver. The main parameters used are listed in Tab. IV.

From both the measured and simulated data, we evaluated the channel gain averaged over angle, which expressed in dB is

$$R_{avg}^{dB}(tx,x,y) = 10\log_{10}\left(\frac{1}{2\pi}\int_0^{2\pi} R(tx,x,y,\phi)d\phi\right), \quad (11)$$

and compared the values obtained from the measurement and simulation to validate the method.

We recall that channel gain is defined as the ratio between the power at the RX antenna port and the power at the TX antenna port. It therefore is also a function of the antenna pattern, that has also been replicated in CST based on anechoic chamber measurement. This definition is necessary as we have no way to compensate for the antenna gain without knowing the elevation pattern of the received signal.

In Fig. 13, we can observe the simulated and measured values of $R_{avg}^{dB}(tx,x,y)$ as a function of the Euclidean distance

---
[3]version 2022

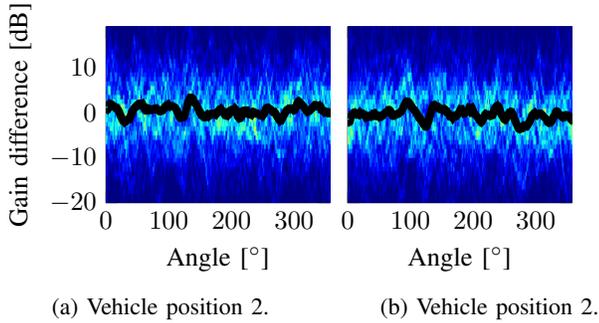

(a) Vehicle position 2.    (b) Vehicle position 2.

Fig. 11: Statistics of the vehicle effect with nonuniform stacking and TX2.

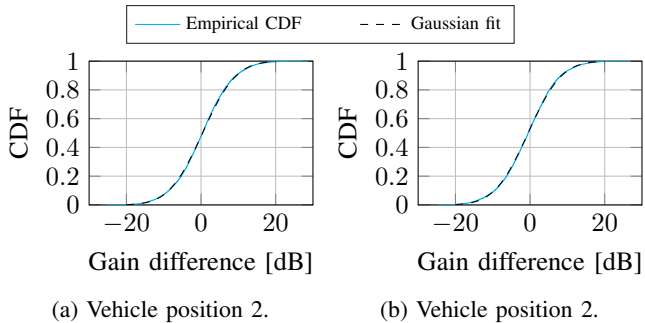

(a) Vehicle position 2.    (b) Vehicle position 2.

Fig. 12: Angle independent statistics of the vehicle effect with nonuniform stacking and TX2.

|  | $\mu$ | $\sigma$ |
|---|---|---|
| Vehicle position 1 | 0.40 | 7.09 |
| Vehicle position 2 | −0.54 | 7.07 |

TABLE III: Channel gain difference Gaussian approximation parameters for the nonuniform stacking and TX2.

| Parameter | Value |
|---|---|
| number of intersections | 5 |
| Ray spacing | 5 $\lambda$ |
| Adaptive ray sampling | yes |
| Maximum ray distance | 10 $\lambda$ |
| Minimum ray distance | 0.2 $\lambda$ |

TABLE IV: Main simulation parameters.

$\bar{D}$ between the TX and RX. As discussed in the previous sections, the multipath is extremely rich, which makes the realization of the channel gain very dependent on fading. As fading changes on the scale of the wavelength, which in this case is a few millimeters, and the positioning of the receiver was not accurate to this scale, we do not expect a match between the measured and simulated power. Therefore we also include a log-linear fit of the data with the equation

$$R_{fit}^{dB}(\bar{D}) = 10n \log_{10}(\bar{D}) + R_0 \qquad (12)$$

The parameters of the fitting lines can be found in Tab. V. Here we can see that the prediction matches the measurements within around 3dB in intercept and 10% in slope, and they agree well within the confidence interval. Based on this result, we consider the simulated channel gain as a good approximation of reality and use it for further studies.

Fig. 14 shows the simulated channel gain when the receiver horn antenna is replaced with an omnidirectional antenna. Comparing the simulated channel gain with the free space path loss associated with the Euclidean distance between the TX and RX, we observe that the difference in channel gain is lower than 10dB for all measured positions. This clearly shows that the highly reflective environment can overcome the limitations of NLoS communication for mmWave. However, the slope, found in Tab. VI, is significantly higher than free space (i.e. $n = 2$ is well outside the 95% confidence interval).

Comparing the results with the horn antenna measurements, which have a slope smaller than 2, we can infer that the further you move from the TX, the more the efficient the horn antenna is. This suggests that, further from the TX, more energy concentrates in the horizontal plane, within the main elevation lobe of the horn. This is also in line with the geometry of the system, as the angle of incidence of the wavefront with the top of the canyon gets shallower at larger distances.

Based on this data, we can estimate the coverage of a typical 5G Base Station (BS) in such environment. To do so, we make the following assumptions:

1) A BS with a transmitting power of 28dBm per polarization per panel, and an antenna gain of 23dBi, for a total EIRP of 51dBm.
2) A shadow-fading margin of 10dB.
3) A bandwidth of 400MHz with at a temperature of 300K and a receiver noise figure of 10$dB$, resulting in a noise floor of $-77.8dBm$.
4) A minimum spectral efficiency of 2 bit/s/Hz (4 bit/s/Hz with dual polarization, resulting in 1.6Gbps with the 400MHz bandwidth), for which we require an Signal to Noise Ratio (SNR) of 8dB [20].
5) A UE with an omnidirectional antenna.

With these assumptions, we have that the maximum allowable path loss is $51 + 77.8 - 8 - 10 = 110.8$dB. We consider the joint fit for the omnidirectional antenna, with a slope of 4.09 and an intercept of $-23.4$. These values result in a channel gain of $-110.8$dB around 137m.

## VII. PROPAGATION MODEL

The observations on the angular spectrum suggests that the energy propagates in a very chaotic and complex manner in the horizontal plane after entering the canyon, therefore we would expect that the amount of energy received by the UE is only dependent on the amount of energy entering the canyon and the vertical propagation inside it.

To test this hypothesis, we compute the received power that we would receive under this assumption. We call $\nu$ the fraction of the energy entering the canyon that actually reaches the receiver. Approximating the signal impinging on the top of the canyon as a plane wave with Poynting vector $S$, the power entering the canyon is proportional to the area of the top opening of the canyon projected on a plane orthogonal to the wave vector of the plane wave. In particular, if we consider the energy entering in a section of the canyon of length $L$

| Configuration | $n$ | $R_0$ [dB] | RMSE[dB] |
|---|---|---|---|
| Measured Uniform | $-1.99 \pm 0.29$ | $-63 \pm 5.67$ | 1.6 |
| Measured Non uniform | $-1.543 \pm 0.33$ | $-67.8 \pm 6.5$ | 1.9 |
| Measured Aggregated | $-2.18 \pm 0.34$ | $-57.4 \pm 6.7$ | 2.71 |
| Simulated Uniform | $-1.88 \pm 0.37$ | $-64 \pm 7.3$ | 1.9 |
| Simulated Non uniform | $-1.78 \pm 0.43$ | $-66.4 \pm 8.7$ | 2.3 |
| Simulated Aggregated | $-1.82 \pm 0.29$ | $-65.3 \pm 5.7$ | 2.1 |

TABLE V: Line fit parameters (95% confidence interval).

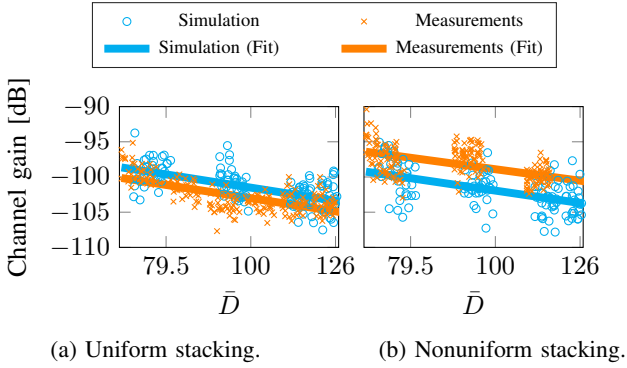

(a) Uniform stacking.   (b) Nonuniform stacking.

Fig. 13: Measured and simulated channel gain.

| | Simulated | | |
|---|---|---|---|
| Stacking | $n$ | $R_0$ [dB] | RMSE[dB] |
| Uniform | $-4.53 \pm 0.29$ | $-16.2 \pm 5.8$ | 2.27 |
| Non uniform | $-3.67 \pm 0.28$ | $-30.3 \pm 5.5$ | 2.2 |
| Aggregate | $-4.09 \pm 0.25$ | $-23.4 \pm 4.9$ | 2.73 |

TABLE VI: Line fit parameters for omnidirectional antenna (95% confidence interval).

and we call the projected canyon aperture $A$, as depicted in Figs. 15 and 16, the received power would be

$$P_{RX} = \nu L A |S|. \quad (13)$$

Using the results in Appendix A, we can rewrite the expression as

$$P_{RX} \propto \frac{\nu L h d}{D^3} = \underbrace{D\psi}_{L} \underbrace{\frac{1}{D^2}}_{\nu} \underbrace{\frac{hd}{D^3}}_{A|S|} = \frac{\psi h d}{D^4}. \quad (14)$$

Notably, the numerator $\psi h d$ is constant. This predicts that the received power decreases with the fourth power of the distance, compared to the free space model which decreases with the second power. This model already matches quite well the observation of Fig. 14 and Tab. VI.

However, to further verify the correctness of the model we fit the data with a slope of $n = 4$, as predicted by (14). Fig. 17 shows such model fit. It is clear that the fit shown is very similar to the one depicted in Fig. 14. Moreover, Tab. VII shows the fit parameters and the Root Mean Square Error (RMSE) obtained with $n = 4$. Clearly, the RMSE obtained has a negligible difference with the one shown in Tab. VI, showing that assuming $n = 4$ does not degrade the fit significantly.

We note that the relevance of this model is not mainly in the ability of predicting the slope, as much as it is in the insight it gives on the propagation mechanisms. Moreover, given the range of distances available, the RMSE is not very sensitive to slope changes. Thus, we can only claim that the data is

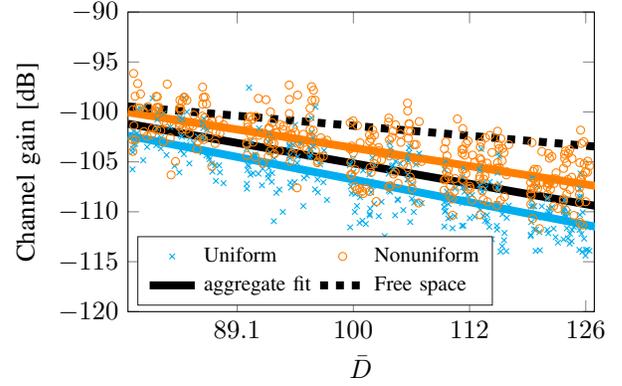

Fig. 14: Simulated channel gain with omnidirectional antennas.

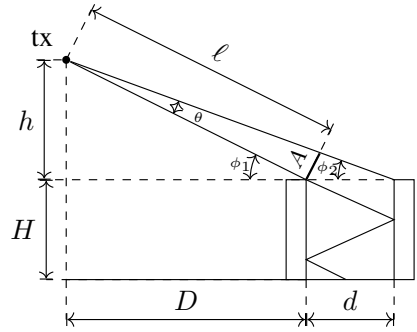

Fig. 15: Container canyon propagation model (side view).

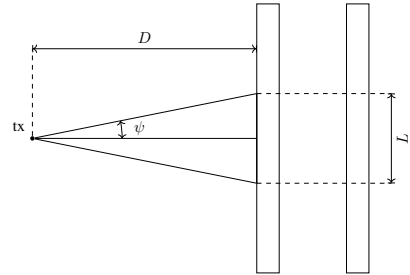

Fig. 16: Container canyon propagation model (top view).

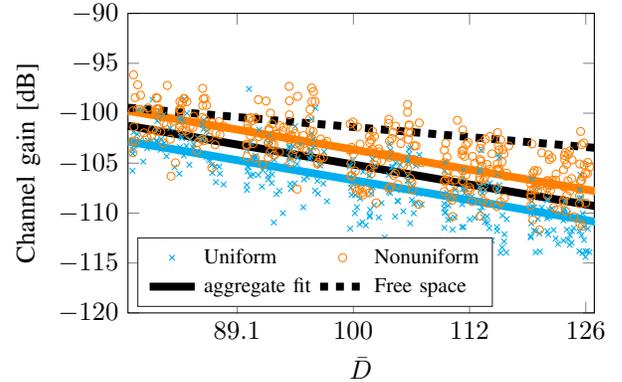

Fig. 17: Channel gain model fit.

consistent with the hypothesis that the received power only depends on the vertical propagation, but needs further evidence

| Stacking | Simulated $R_0$ [dB] | RMSE[dB] |
|---|---|---|
| Uniform | $-26.7 \pm 0.22$ | 2.3 |
| Non uniform | $-23.63 \pm 0.21$ | 2.22 |
| Aggregate | $-25.13 \pm 0.19$ | 2.73 |

TABLE VII: Model fit parameters (95% confidence interval).

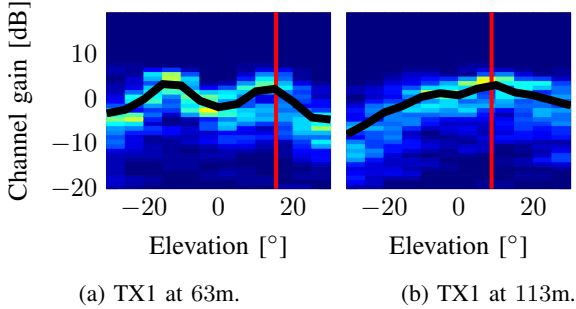

(a) TX1 at 63m.  (b) TX1 at 113m.

Fig. 18: Elevation spectrum for the nonuniform configuration.

to reinforce the claim, which will be presented in section VIII.

## VIII. ELEVATION SPECTRUM

Fig. 18 shows the relation between the channel gain and the elevation angle of the RX antenna. It was realized simulating the channel with the horn antenna keeping the azimuth angle fixed at $\phi = 90°$, i.e. pointed towards the side of the canyon where the transmitter is located, and rotating it to different elevation angles. The simulation is repeated for two positions of TX1, the first at 63m and the second at 113m. In red, we show the related angle $\phi_1$ computed according to (18), which is of $15.4°$ and $8.8°$ respectively.

Here we can clearly see that there is a dependence of the elevation angle of arrival of the distance between the transmitter and the canyon. In particular, we can observe how the maximum of the channel gain corresponds with the geometric angle, thus reinforcing the propagation hypotheses stated above. This fact also explains the difference in slope observed between the measurements and simulations with the horn antenna (Tab. V) and the simulations with the omni antenna (Tab. VI). In fact, in the former case the slope results lower because the further we move, the shallower the elevation angle is, which means that the gain of the antenna is better exploited at larger distances.

## IX. CONCLUSION

We have conducted measurements to characterize the propagation of mmWaves in a container canyon in a port environment with a narrow-band channel sounder. These measurements were utilized to evaluate the angular spectrum and spatial correlation inside the canyon. It was observed that energy was received uniformly from all directions except the ends of the canyon. In addition, for a given beam direction, the channel becomes completely uncorrelated over short distances of the order of 10 cm inside the canyon implying that the 'best beam' direction over a given bandwidth is short lived. From both of these observations, we can conclude that the beamforming gain from frequency-flat azimuthal beamforming is minimal.

Given that the measurements were obtained with a highly directional horn antenna, to determine the pathloss and elevation spectrum, measurements with an isotropic antenna would be needed. We validated a simulation setup with these measurements and observed that the gap between the simulated and measured results were small. We then replaced the directional antennas with omnidirectional ones to obtain simulated values of the pathloss and elevation spectrum. We observed that the difference in channel gain between the omnidirectional receive antenna and the free-space path loss is less than 10 dB for all the measured positions, indicating that the highly reflective surface of the containers can compensate for the path loss at mmWave and provide rich multi-path.

Lastly, we provide an explanation for the mechanism of wave propagation that determines the path loss inside the canyon. We hypothesize that the received energy is only dependent on the energy entering the canyon and the vertical propagation inside it. We validate this hypothesis with the simulated data.

## APPENDIX A

Since there are typically no objects above the canyon level, $|S|$ can be computed as the Free Space Path Loss (FSPL) between the transmitter and the canyon, which in this case is proportional to

$$|S| \propto \ell^{-2} = \left(\sqrt{h^2 + D^2}\right)^{-2}. \quad (15)$$

Here we ignore the proportionality constant, as we can fit the model to the data later. Assuming $\theta$ is small, the aperture $A$ can be approximated as

$$A \approx \ell \sin(\theta) = \left(\sqrt{h^2 + D^2}\right) \sin(\theta). \quad (16)$$

We note that

$$\theta = \phi_1 - \phi_2; \quad (17)$$

$$\phi_1 = \tan^{-1}\left(\frac{h}{D}\right); \quad (18)$$

$$\phi_2 = \tan^{-1}\left(\frac{h}{D+d}\right). \quad (19)$$

Therefore, we can write

$$\theta = \tan^{-1}\left(\frac{h}{D}\right) - \tan^{-1}\left(\frac{h}{D+d}\right) \quad (20)$$

and

$$A \approx \ell \sin(\theta) = \left(\sqrt{h^2 + D^2}\right) \sin\left(\tan^{-1}\left(\frac{h}{D}\right) - \tan^{-1}\left(\frac{h}{D+d}\right)\right). \quad (21)$$

We can now rewrite the received power as

$$P_{RX} = \nu LA|S| \tag{22}$$

$$\approx \nu L \left(\sqrt{h^2 + D^2}\right) \times$$
$$\sin\left(\tan^{-1}\left(\frac{h}{D}\right) - \tan^{-1}\left(\frac{h}{D+d}\right)\right) \left(\sqrt{h^2 + D^2}\right)^{-2} \tag{23}$$

$$= \nu L \left(\sqrt{h^2 + D^2}\right)^{-1} \times$$
$$\sin\left(\tan^{-1}\left(\frac{h}{D}\right) - \tan^{-1}\left(\frac{h}{D+d}\right)\right). \tag{24}$$

We can now perform some small angle approximations to further simplify the expression. We assume $D \gg h$, i.e. we consider the case where the transmitter is far from the canyon, which is the most relevant and challenging case as it is the one representing the cell edge case. Thanks to this assumption, we can say that also $\phi_1$ and $\phi_2$ are small, and use a first order approximation of the arctangent in (18) and (19). Therefore, we write

$$\phi_1 \approx \frac{h}{D} \quad \text{and} \quad \phi_2 \approx \frac{h}{D+d}. \tag{25}$$

This simplifies the received power to

$$P_{RX} \approx \nu L \left(\sqrt{h^2 + D^2}\right)^{-1} \sin\left(\frac{h}{D} - \frac{h}{D+d}\right)$$
$$= \nu L \left(\sqrt{h^2 + D^2}\right)^{-1} \sin\left(\frac{hd}{D^2 + dD}\right). \tag{26}$$

Furthermore, we have that $h^2 + D^2 \approx D^2$, further simplifying the expression to:

$$P_{RX} \approx \frac{\nu L}{D} \sin\left(\frac{hd}{D^2 + dD}\right). \tag{27}$$

Under the assumption of far transmitter, it also holds that $D \gg d$, hence we can approximate $D^2 + dD$ with $D^2$, obtaining

$$P_{RX} \approx \frac{\nu L}{D} \sin\left(\frac{hd}{D^2}\right). \tag{28}$$

Finally, by the assumptions above, we note that $D^2 \gg hd$, and therefore $\frac{hd}{D^2}$ is small, so we can use the first order approximation of the sine to obtain the final expression

$$P_{RX} \approx \frac{\nu L}{D} \frac{hd}{D^2} = \frac{\nu Lhd}{D^3}. \tag{29}$$

The value of $L$ can be computed assuming that the energy reaches the receiver only if it enters the canyon at a shallow azimuth angle. Fig. 16 illustrates the geometry of the system.

In this figure, we can easily see that, calling the maximum azimuthal angle $\psi$ and assuming that only the energy within the wedge reaches the receiver, the length of the canyon section accepting energy can be expressed as

$$L = D\sin(\psi) \approx D\psi. \tag{30}$$

Let us now consider the value of $\nu$. The length of the path followed by the signal inside the canyon corresponds to the length of $\ell'$ in Fig. 19.

By triangles similarity we can write

$$\frac{h}{D} = \frac{h'}{D'} \Rightarrow D' = \frac{h'D}{h}. \tag{31}$$

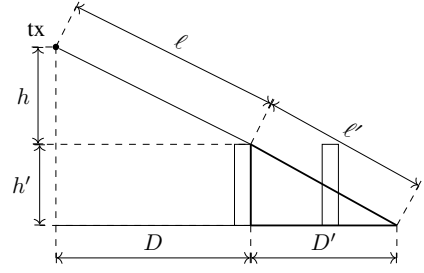

Fig. 19: Propagation inside the canyon.

We can then write

$$\ell' = \sqrt{(h')^2 + (D')^2} = \sqrt{(h')^2 + \left(\frac{h'D}{h}\right)^2}. \tag{32}$$

By the assumption that $D$ is large we can approximate it with

$$\ell' \approx \sqrt{(D')^2} = \sqrt{\left(\frac{h'D}{h}\right)^2} = \frac{h'D}{h} \tag{33}$$

and, noting that by the plane wave approximation the energy is equally spread in the incident area, write

$$\nu \propto \frac{1}{(\ell')^2} = \left(\frac{h}{h'D}\right)^2 \propto \frac{1}{D^2}. \tag{34}$$

ACKNOWLEDGMENT

The authors would like to thank Cargotec Kalmar for providing access to their facilities and constructing the experiment site.

REFERENCES

[1] L. Fanari, E. Iradier, I. Bilbao, R. Cabrera, J. Montalban, P. Angueira, O. Seijo, and I. Val, "A survey on fec techniques for industrial wireless communications," *IEEE Open Journal of the Industrial Electronics Society*, vol. 3, pp. 674–699, 2022.
[2] A. Ahmadi, M. Moradi, C. Cherifi, V. Cheutet, and Y. Ouzrour, "Wireless connectivity of cps for smart manufacturing: A survey," in *2018 12th International Conference on Software, Knowledge, Information Management and Applications (SKIMA)*, 2018, pp. 1–8.
[3] P. Angueira, I. a. Val, J. Montalbán, O. Seijo, E. Iradier, P. S. Fontaneda, L. Fanari, and A. Arriola, "A survey of physical layer techniques for secure wireless communications in industry," *IEEE Communications Surveys and Tutorials*, vol. 24, no. 2, pp. 810–838, 2022.
[4] M. Cheffena, "Industrial wireless communications over the millimeter wave spectrum: Opportunities and challenges," *IEEE Communications Magazine*, vol. 54, no. 9, pp. 66–72, 2016.
[5] J. Mazgula, J. Sapis, U. S. Hashmi, and H. Viswanathan, "Ultra reliable low latency communications in mmwave for factory floor automation," *Journal of the Indian Institute of Science*, vol. 100, pp. 303–314, 2020.
[6] M. A. Uusitalo, H. Viswanathan, H. Kokkoniemi-Tarkkanen, A. Grudnitsky, M. Moisio, T. Härkönen, P. Yli-Paunu, S. Horsmanheimo, and D. Samardzija, "Ultra-reliable and low-latency 5g systems for port automation," *IEEE Communications Magazine*, vol. 59, no. 8, pp. 114–120, 2021.
[7] J. Mietzner, R. Schober, L. Lampe, W. H. Gerstacker, and P. A. Hoeher, "Multiple-antenna techniques for wireless communications - a comprehensive literature survey," *IEEE Communications Surveys and Tutorials*, vol. 11, no. 2, pp. 87–105, 2009.
[8] S. Kutty and D. Sen, "Beamforming for millimeter wave communications: An inclusive survey," *IEEE Communications Surveys and Tutorials*, vol. 18, no. 2, pp. 949–973, 2016.
[9] I. Ahmed, H. Khammari, A. Shahid, A. Musa, K. S. Kim, E. De Poorter, and I. Moerman, "A survey on hybrid beamforming techniques in 5g: Architecture and system model perspectives," *IEEE Communications Surveys and Tutorials*, vol. 20, no. 4, pp. 3060–3097, 2018.

[10] J. Hao and H. Liu, "Fast spatial correlation acquisition for hybrid precoding using sequential compressive sensing," in *2018 15th IEEE Annual Consumer Communications Networking Conference (CCNC)*, 2018, pp. 1–6.

[11] S. Chen, Q. Gao, R. Chen, H. Li, S. Sun, and Z. Liu, "A csi acquisition approach for mmwave massive mimo," *China Communications*, vol. 16, no. 9, pp. 1–14, 2019.

[12] D. Chizhik, J. Du, R. Feick, M. Rodriguez, G. Castro, and R. A. Valenzuela, "Path loss and directional gain measurements at 28 GHz for non-line-of-sight coverage of indoors with corridors," *IEEE Transactions on Antennas and Propagation*, vol. 68, no. 6, pp. 4820–4830, 2020.

[13] Q. Spencer, B. Jeffs, M. Jensen, and A. Swindlehurst, "Modeling the statistical time and angle of arrival characteristics of an indoor multipath channel," *IEEE Journal on Selected Areas in Communications*, vol. 18, no. 3, pp. 347–360, 2000.

[14] D. Chizhik, J. Du, R. A. Valenzuela, D. Samardzija, S. Kucera, D. Kozlov, R. Fuchs, J. Otterbach, J. Koppenborg, P. Baracca, M. Doll, I. Rodriguez, R. Feick, and M. Rodriguez, "Directional measurements and propagation models at 28 ghz for reliable factory coverage," *IEEE Transactions on Antennas and Propagation*, vol. 70, no. 10, pp. 9596–9606, 2022.

[15] D. Chizhik, J. Du, R. A. Valenzuela, J. Otterbach, R. Fuchs, and J. Koppenborg, "Path loss and directional gain measurements at 28 GHz for factory automation," in *2019 IEEE International Symposium on Antennas and Propagation and USNC-URSI Radio Science Meeting*, 2019, pp. 2143–2144.

[16] D. Solomitckii, M. Allén, D. Yolchyan, H. Hovsepyan, M. Valkama, and Y. Koucheryavy, "Millimeter-wave channel measurements at 28 GHz in digital fabrication facilities," in *2019 16th International Symposium on Wireless Communication Systems (ISWCS)*, 2019, pp. 548–552.

[17] D. Solomitckii, A. Orsino, S. Andreev, Y. Koucheryavy, and M. Valkama, "Characterization of mmWave channel properties at 28 and 60 GHz in factory automation deployments," in *In proc. IEEE Wireless Communications and Networking Conference (WCNC), 2018*, 2018, pp. 1–6.

[18] C. Cano, G. H. Sim, A. Asadi, and X. Vilajosana, "A channel measurement campaign for mmwave communication in industrial settings," *IEEE Transactions on Wireless Communications*, vol. 20, no. 1, pp. 299–315, 2021.

[19] M. Ghaddar, I. Ben Mabrouk, M. Nedil, K. Hettak, and L. Talbi, "Deterministic modeling of 5G millimeter-wave communication in an underground mine tunnel," *IEEE Access*, vol. 7, pp. 116 519–116 528, 2019.

[20] R. Kovalchukov, D. Moltchanov, Y. Gaidamaka, and E. Bobrikova, "An accurate approximation of resource request distributions in millimeter wave 3gpp new radio systems," 08 2019.